\documentclass[10pt]{article}

\usepackage{amsmath,amssymb,amsfonts}
\usepackage{graphicx}

\newcommand{\myfig}[3]{
\begin{figure}
\centering
\includegraphics[width=#2cm]{#1}\caption{#3}\label{#1}
\end{figure}
}

\begin{document}

\def\CH{{\cal H}}
\def\CO{{\cal O}}
\newcommand{\be}{\begin{equation}}
\newcommand{\ee}{\end{equation}}
\newcommand{\bea}{\begin{eqnarray}}
\newcommand{\eea}{\end{eqnarray}}
\def\bes{\begin{eqnarray}}
\def\ees{\end{eqnarray}}
\newcommand{\pa}{\partial}
\newcommand{\eref}[1]{(\ref{#1})}
\def\tr{tr}
\newcommand{\BC}{\mathbb{C}}
\newcommand{\BR}{\mathbb{R}}
\newcommand{\BZ}{\mathbb{Z}}
\newcommand{\extd}{{\rm d}}
\newcommand{\ym}[1]{YM$_{#1+1}$}

\renewcommand{\thepage}{\arabic{page}}
\setcounter{page}{1}

\rightline{hep-th/0604184}
\rightline{ILL-(TH)-06-04}
\rightline{VPI-IPPAP-06-05}

\vskip 0.75 cm
\renewcommand{\thefootnote}{\fnsymbol{footnote}}
\centerline{\Large \bf Towards a solution of pure Yang-Mills theory
in $3+1$ dimensions}
\vskip 0.75 cm

\centerline{{\bf
Laurent Freidel,${}^{1,2}$\footnote{lfreidel@perimeterinstitute.ca}
Robert G. Leigh${}^{3}$\footnote{rgleigh@uiuc.edu},
and Djordje Minic${}^{4}$\footnote{dminic@vt.edu}
}}
\vskip .5cm
\centerline{${}^1$\it Perimeter Institute for Theoretical Physics,}
\centerline{\it 31 Caroline St., Waterloo, ON, N2L 2Y5, Canada. }
\vskip .5cm
\centerline{${}^2$\it Laboratoire de Physique,
\'Ecole Normale Sup\'erieure de Lyon}
\centerline{\it 46 All{\'e}e d'Italie, 69364 Lyon, Cedex 07, France.}
\vskip .5cm
\centerline{${}^3$\it Department of Physics,
University of Illinois at Urbana-Champaign}
\centerline{\it 1110 West Green Street, Urbana, IL 61801-3080, USA}
\vskip .5cm
\centerline{${}^4$\it Institute for Particle Physics and Astrophysics,}
\centerline{\it Department of Physics, Virginia Tech}
\centerline{\it Blacksburg, VA 24061, USA}
\vskip .5cm

\setcounter{footnote}{0}
\renewcommand{\thefootnote}{\arabic{footnote}}

\begin{abstract}
We discuss an analytic approach towards the solution of pure Yang-Mills
theory in $3+1$ dimensional spacetime. The approach is based
on the use of local gauge invariant variables in the Schr\"odinger
representation and the large $N$, planar limit. In particular, within  
this
approach we point out unexpected parallels between pure Yang-Mills
theory in $2+1$ and $3+1$ dimensions. The most important parallel shows
up in the analysis of the ground state wave-functional especially in
view of the numerical similarity of the existing large N lattice
simulations of the spectra of $2+1$ and $3+1$ Yang Mills theories.
\end{abstract}
\newpage

\section{Introduction}

Recently, new analytical results pertaining to the spectrum of $2+1$
dimensional Yang-Mills theory (\ym{2}), that are in excellent agreement
with the lattice data in the planar limit \cite{teper}, have been
derived in \cite{prl}. These analytic results regarding the mass gap,
string tension and the glueball spectrum resulted from a determination
of the ground state wave-functional in the planar limit \cite{thooft}.
The analytic calculations in \ym{2} were based, as in \cite{robme}, on
the remarkable work of Karabali and Nair \cite{knair}.

Because in $2+1$ dimensions space is two dimensional, one can work in a
complex basis in the Hamiltonian picture. One
might wonder whether a similar success can be achieved in
the more realistic case of a $3+1$ dimensional Yang-Mills theory (\ym 
{3}).
As we will explain in this letter, a formalism does exist which closely
parallels that of KN. The ingredients of this formalism were introduced
long ago by I. Bars \cite{bars} in his work on local gauge invariant
`corner variables'. (Other related work can be found in \cite{other}.)

In this rather programmatic letter, we outline and extend the general
formalism, and explain the physical interpretation and use of corner
variables both in \ym{2} and \ym{3}. In particular, in the context of
\ym{2}, the corner variables are appropriate to a real coordinate basis
as opposed to the complex basis of KN. Also, as pointed out by Bars in
\cite{bars}, the corner variables can be also used in a covariant
Lagrangian picture, which is an obvious additional attractive feature of
the formalism. In the Hamiltonian picture the use of corner variables
also reveals unexpected parallels between pure Yang-Mills theory in $2+1$
and $3+1$ dimensions. We concentrate here on the discussion of the
vacuum wave-functional in the Hamiltonian picture,
especially in light of the large N lattice simulations.

The themes outlined in this paper are explored in more detail in a
companion paper \cite {L} and in \cite{4dpapers}. The organization of  
this
letter is as follows: in Section 2 we discuss the formalism of corner
variables in view of our recent work on pure Yang Mills theory in 2+1
dimensions \cite{prl}. Then we turn to the more dynamically relevant
issues in Section 3. Section 4 is devoted to the general analysis of the
structure of the vacuum wave-functional. We conclude the paper with a
short outline of some of our current work in progress.

\section{The Formalism of Corner Variables}

The key insight we start from is that the KN variables \cite{knair} in
the $2+1$ dimensional setting are related to line integrals in from
infinity to a point $x$. The analogous variables in $3+1$ were
introduced by Bars and we call them $M_{i}(x)$ where $M_i(x)$ are
unitary matrices.\footnote{We use notation similar to that of KN and
warn the reader of notation difference compared to Bars.} They satisfy
the defining equation
\begin{equation}
A_{i} = - \partial_{i} M_{i}M_{i}^{-1} \ \ \ ({\rm no\ sum\ on}\ i)
\end{equation}
In other words
\begin{equation} M_i(x) = Pexp [-\int_{-\infty}^x A]
\end{equation}
where the integral is a straight spatial contour for fixed $x^j$, $i\neq
j$. The interesting thing about these variables is that the natural
lattice formulation {\it parallels} the continuum formulation. The above
formulation is appropriate to the Hamiltonian formalism, if $j$ is a
spatial index. The translation to 2+1 KN variables is then just
\begin{equation}
M_z=M,\ \ \ \ \ M_{\bar z}={M^{\dagger}}^{-1}
\end{equation}
Note that if we had used a real coordinate basis, then we would have
had a pair of (unrelated) {\it unitary} matrices $M_1$, $M_2$.
Next, define the corner variables
\begin{equation}
H_{ij}=M_i^{-1}M_j
\end{equation}
\myfig{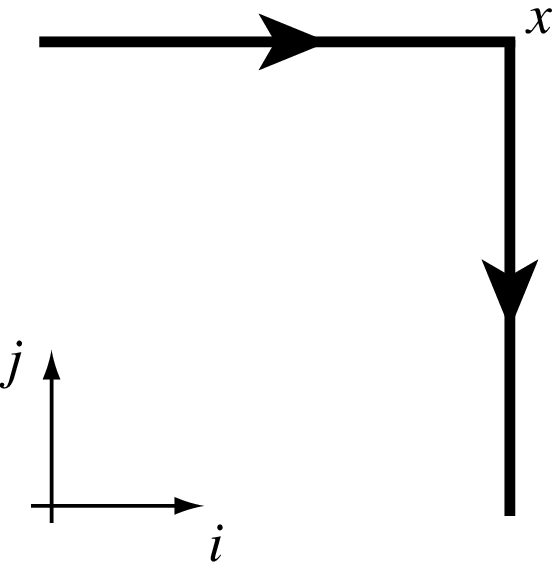}{3}{The `corner variable' $H_{ij}(x)$.}
Note that $H_{jj}=1$ and $H_{ji}=H_{ij}^{-1}$ -- this just means
traversing the corner in the opposite direction is precisely the inverse
group element. The $H_{ij}$ are {\it unitary} in a real coordinate  
basis.
There is also a constraint (here written for $3+1$ --- it is trivial in
$2+1$)
\begin{equation}\label{constraint}
H_{ij}H_{jk}H_{ki}=1
\end{equation}

Another description is in terms of a ``semi-complex'' coordinate basis
$\{ u,z,\bar z\}$; then one may parameterize as
\begin{eqnarray}
H_{uz}=H,\quad
H_{\bar z u}=H^\dagger,\quad
H_{\bar z z} = H^\dagger H.
\end{eqnarray}
For example, one could use the notation $M_z=M$, $M_{\bar z}={M^ 
{\dagger}}^{-1}$, $M_u^\dagger M_u=1$, with $H=M_u^{\dagger}M$. The  
constraint
takes the form $H_{\bar zu}H_{uz}H_{z\bar z}=1$. In other words, there
is in $D=3$ in the semi-complex coordinate basis a {\it complex}
$H$-field (compared to a Hermitian field in $D=2$); thus there are twice
as many degrees of freedom, as expected. Returning to the general
notation, gauge transformations act as
\begin{equation}
   M_j\mapsto g M_j
\end{equation}
so the $H_{ij}$ are gauge invariant.

The formalism of Bars also contains the notion of (or a generalization
of) holomorphic invariance (which is familiar from the $2+1$ KN
setting). Indeed, note that one may introduce `currents'
\begin{equation}
J_{ij}=(\partial_j H_{ij}) H^{-1}_{ij}
\end{equation}
(in 2+1, $J\sim J_{\bar zz}$ and $J^\dagger\sim -J_{z\bar z}$.) The  
extra symmetry acts as
\begin{equation}
M_i\mapsto M_i\ h^{-1}_i(x^j),\ \ \ \ j\neq i
\end{equation}
(this apparently, as far as we
can tell, was not pointed out in \cite{bars}.) The condition $j\neq i 
$ on the function $h_i$ is the analogue of
holomorphy. This leaves the gauge fields invariant, and one finds
\begin{equation}
H_{ij}\mapsto h_iH_{ij}h_j^{-1}
\end{equation}
In the semi-complex basis, we would have $M_u\to M_u\ h^{-1}_u(z,\bar
z)$, $M_z\to M_z\ h^{\dagger}(u,\bar z)$ and so $H\to h_u(z,\bar z)H
h^\dagger (u,\bar z)$, $H^\dagger\to h(u,z)H^\dagger h_u^{-1}(z,\bar
z)$. The $J_{ij}$ transform as connections
\begin{equation}
J_{ij}\mapsto h_i J_{ij}h_i^{-1}+\partial_j h_i\ h_i^{-1}
\end{equation}

In the real coordinate basis, it appears that there are six currents
that are apparently distinct. However, there is a 'reality' condition on
their derivatives of the form
\begin{equation}\label{dJ2}
\partial_i J_{ij}=-H_{ij}(\partial_j J_{ji})H_{ij}^{-1}
\end{equation}
(in $D=2$, there is a similar relation which reads $\bar\partial J=
H(\partial J^\dagger) H^{-1} $.) By defining $\bar
J_{ij}=-H_{ij}J_{ji}H_{ij}^{-1}$, we may rewrite this as
\begin{equation}
\partial_i J_{ij}=\partial_j\bar J_{ij}-[J_{ij},\bar J_{ij}]
\end{equation}
and so there are covariant derivatives $D_{ij}=\partial_j-J_{ij}$. These
currents are related to the magnetic field $F_{ij}=
\partial_{i}A_{j}-\partial_{j}A_{i}+ [A_{i},A_{j}]$ by
\be
\partial_i J_{ij}= - M_{i}^{{-1}}F_{ij}M_{i}.
\ee
As a short form then, we will denote
\be
B_{i-1}=\pa_i J_{i,i+1}\ \ \ ({\rm cyclic}).
\ee
These fields transform homogeneously under the `holomorphic'
symmetry
\be
B_{i-1}\mapsto h_i B_{i-1} h_i^{-1}
\ee
Note that there are only really two of these fields, because, solving
the constraint (\ref{constraint}), there are only two independent
$H_{ij}$'s, say $H_{12}$ and $H_{13}$.


Note that the covariant derivatives can be written in terms of the
usual derivative and $M_{i}$ as
\be\label{covd}
\nabla_{i}=\partial_{i}+ A_{i}= M_{i}\partial_{i} M_{i}^{-1}.
\ee

Finally we note some notations and conventions used in this paper. The
connection is expanded in terms of {\it anti-hermitean} generators
$T_{a}$, $A_{i}= A_{i}^{a} T_{a}$ satisfying the algebra $[T_{a},
T_{b}]= f_{ab}{}^{c}T_{c}$.  Also, we denote by $\tr$ the trace in the
fundamental representation, so that $\tr (1)=N$. This trace is
normalized as $-2 \tr(T_{a}T_{b})= \delta_{ab}$. In the adjoint
representation the generators are given by $(T^{a})_ {bc}= -f_{abc}$,
where indices are raised or lowered with the metric $\delta_{ab}$ and
the trace in the adjoint is denoted by $\tr_{ad}$. A group element $M$
is represented in the adjoint by $M_{ab}= -2\tr(T_{a}MT_{b}M^{-1})$,
clearly we have $(M^{-1})_{ab}= M_{ba}$ and also $MT_{b}M^{-1}= T_{a}
M^{a}{}_{b}$.

We will need a variety of Green's functions. We denote by $G_{i}(x,y)$
the inverse of $\partial_{i}$, such that $\partial_{i}^x G_{i}(x,y)=
\delta^{(D)}(x,y)$ (no sum). Since $\partial_{i}$ admits zero modes, its
inverse is not uniquely defined and we will work with the explicit
choice $G_{1}(x)=  \theta(x_{1})\delta({x_{2}})\delta({x_{3}})$ and
$G_{i} (x,y)\equiv G_{i}(x-y)$ (which is not antisymmetric). This choice
is not arbitrary, it is the unique choice consistent with the definition
of the variable $M_{i}$ as an ordered exponential. Indeed we can write
\be \label{ordprop}
M_{i}(x) = \sum_{n}(-1)^n \int \extd y (G_{i}A_{i})^n(x,y),
\ee
where $(G_{i}A_{i})^2(x,y)= \int \extd z G_{i}(x,z)A_{i}(z)G_{i}(z,y)
A_{i}(y) $ .

\section{Further Technical Details}

The Yang-Mills action is taken to be $S_{YM}= \frac1{2g^2} \int\tr
(F_{\mu\nu}F^{\mu\nu})$ and the Hamiltonian in the Yang-Mills variables
is
\be
{\mathcal{H}}  = \sum_{i,a}
\left\{ -\frac{g^2}{2} \left(\frac{\delta}{\delta A_{i}^{a}}\right)^2
+ \frac{1}{2g^2} (F_{i}^{a})^2\right\}
\ee
with $F_{i}^{a}=\frac12 \epsilon_{ijk}F_{jk}^{a}$ and should be
supplemented by the gauss law constraint $\nabla_{i}
\frac{\delta}{\delta A_{i}^{a}} =0$. The wave-functionals in the
Schr\"odinger representation of pure Yang-Mills are gauge invariant
functionals of $A_{i}$ and the scalar product is given by
\be
||\Psi||^2 = \int_{\cal{A}/\cal{G}} D\mu(A)\ \bar{\Psi}[A] \Psi[A],
\ee
Where the integral is over the space of gauge connections modulo gauge
transformations, $D\mu(A)= \frac{DA}{\mathrm{Vol(\cal{G})}}=
\frac{\prod_{i}DA_{i}} {\mathrm{Vol(\cal{G})}}$. From the previous
section we know that we can equivalently describe gauge invariant wave
functionals as `holomorphic' invariant wave-functionals of $H_{ij}$ or
$J_{ij}$. This  change of variables can be explicitly performed both at
the level of the measure and the Hamiltonian  \cite{L}; we recall here
some of these results.

Since $A_{i}= -\partial_{i}M_{i}M_{i}^{-1}$, $\delta A_{i}=
-(\nabla_{i}\delta M_{i}) M_{i}^{-1} $, the change of variables involves
a determinant
\be
e^\Gamma\equiv \mathrm{det}\left(\frac{\delta A_{i}}{\delta M_{i}M_{i}
^{-1}}\right)= \mathrm{det}(\nabla_{1}\nabla_{2}\nabla_{3}).
\ee
The variational derivative of the action is found to be trivial
\be\label{delG}
\frac{\delta \Gamma}{\delta A_{i}^{a}(x)} = \tr_{ad}\left[(\nabla_{i})
^{-1}(x,x)T_{a}\right] =1.
\ee
The scalar product written in term of the Bars variables is  simply
\be
||\Psi||^2 = \int DH_{12} DH_{13}\ \bar{\Psi}[H] \Psi[H] = \int \prod_
{i<j}DH_{{ij}}\ \delta(H_{12}H_{13}H_{23})\ \bar{\Psi}[H] \Psi[H]
\ee
where $DH$ denotes the product over left-right Haar measures on $SU(N)$.
In the last equality we inserted a delta function on the group to
emphasize the symmetric form of the measure under  permutation of
indices. The interpretation of this delta function constraint should be
clear: it is the integrated version of the Bianchi identity
$\epsilon^{ijk} \nabla_{i}F_{jk}=0$ expressed in terms of gauge
invariant observables. One important feature of this scalar product is
the fact that the identity functional $\Psi[H]=1$, which can be can be
viewed as a limit of the identity cylindrical functional, is a
normalisable wave-functional.

The potential term of the Yang-Mills Hamiltonian is easily expressed in
terms of the Bars variables
\be
\mathcal{V} \equiv \frac12\sum_{i,a}(F_{i}^{a})^2 = \frac12\sum_{i,a}
\int (\partial_{i}J_{i, i+1}^{a})^2(x) \extd x.
\ee
and the kinetic term reads
\be\label{T1}
\mathcal{T}= -\frac12 \sum_{a,i} \int\frac{\delta }{\delta A_{i}^{a}(x)}
\frac{\delta}{\delta A_{i}^{a}(x)}
= \frac12\sum_{i}\int \extd y \extd z\
P_{a}^{i}(y) {\Theta}^{ab}_{i}(y,z) P_{b}^{i}(z)
\ee
where $P_{a}^{i}$ are right derivatives on the group
\be
[P^{i}_{a}(x), M_{j}(y)] \equiv
\delta^{j}_{i}\ M_{i}(y) T_{a}\ \delta^{(D)}(x,y)
\ee
and we have introduced the kernel
\be
{\Theta}_{i}^{ab}(y,z) = \delta^{ab}(G_{i}G_{i})(y,z)= \delta^{ab}
\int\extd x\ G_{i}(y,x)G_{i}(x,z).
\ee
It is important to note that the form (\ref{T1}) for the kinetic term is
valid only when $\mathcal{T}$ acts on holomorphic invariant states. This
Hamiltonian can be checked to be hermitian \cite{L}.

Two technical points are in order here. First, in the KN formalism, a
non-trivial Jacobian appeared and is given by the exponent of a gauged
WZW action. This is an artifact of the complex polarization adopted in
the formalism, and it is not a universal feature. In the real basis that
we have discussed here, such a Jacobian does not arise. Indeed, consider
the computation of such a Jacobian determinant, which one can write
conveniently as a fermionic determinant. Depending on the choice of a
chiral (in the complex KN-like polarization) or non-chiral basis (in the
real Bars-like polarization) for the auxiliary adjoint fermions, in the
case of $2+1$ Hamiltonian formulation, one gets the usual anomalous term
represented by the WZW action, or one gets a unit Jacobian,
respectively. Thus one also has a trivial unit Jacobian in $3+1$
dimensions if one works with Bars' corner variables.

Second, there is a delicate issue of holomorphically invariant
regularization procedure (which extends the techniques used by
\cite{knair}). And indeed, in order to make sense of the computation of
the determinant or the hermiticity property of the kinetic term, one
needs a regularisation scheme which preserves all the symmetries. Such a
regularisation scheme is fully described in \cite{L} and is absolutely
crucial for actual computations.

\subsection{Semi-complex coordinates}

One of the most important observations about the Bars' formalism is that
the kinetic terms have the form $\mathcal{T} = \int (G pq)^2$
(where $p$ and $q$
denote the canonical generalized momentum and coordinate, $[q,p] =i$)
with a regularized Green's function $G$.

The significance of this will be appreciated by considering a toy
example with a similar Hamiltonian. Consider a free 1d rotator. The
kinetic term is $ T = - \frac{\delta^2}{\delta \theta^2}. $ Here,
$\theta$ is a periodic variable and the spectrum is discrete, with
eigenvalues $n^2$, and with periodic eigenfunctions $e^{i n \theta}$.
Now if we change variables $\rho = e^{i\theta}$, the kinetic term
becomes homogeneous
\begin{equation}
\mathcal{T} = ( \rho \frac{\delta}{\delta \rho} \rho
\frac{\delta}{\delta \rho} )
\end{equation}
where obviously $\rho$ and its conjugate momentum $\frac{\delta} {\delta
\rho}$ have the canonical commutation relations. So this has a $(qp)^2$
form, as required. Now we can rewrite this using the canonical
commutator as
\begin{equation}
\mathcal{T} = \rho \frac{\delta}{\delta \rho} +
\rho^2 (\frac{\delta}{\delta\rho})^2
\end{equation}
which looks like the usual kinetic term for pure Yang-Mills theory in
terms of complex corner variables as in \cite{knair}. We look at the
action of this kinetic term on $ \rho^n, \quad n=1,2,3... $ which
correspond to the eigenstates $e^{i n \theta}$. The first term gives a
homegeneity factor (the Euler factor) of $n$, which is of course not the
correct eigenvalue! To get the correct eigenvalue $n^2$ we need also the
second term in $\mathcal{T}$\footnote{Note that in the quantum field  
theory setting the
factor of $n^2$ is {\it misleading} as it can be seen in the detailed  
analysis of the
$2+1$ dimensional Yang-Mills theory \cite{prl}. Further discussion is in Section 4.}.

The upshot here is that we see a close parallel between this simple toy
model and the formulation of pure Yang-Mills theory in terms of corner
variables.

In view of this toy example we discuss the complex versus real corner
variables and the form of the corresponding Hamiltonian. The similarity
between the Karabali-Nair Hamiltonian written in terms of complex corner
variables and the collective field theory \cite{collective} has been
noted before \cite{knair}, \cite{robme}, \cite{prl}. Yet the Hamiltonian
written in terms of real variables apparently does not have this form.
In particular, it seems that the homogeneous piece of the Hamiltonian
(as illustrated by the above toy example) is missing. One would have
such a term however, if we write the theory in the  semi-complex basis
in $3+1$ that we discussed in Section 2. Such a term acts homogeneously
on suitable functionals and thus acts as an effective dimension counting
operator. In what follows we will further comment on this issue when we
discuss the constituent picture of glueballs.

More explicity in terms of the semi-complex coordinates from section 2
($u, z, \bar{z}$), the gauge invariant information is encoded in a
complex $H \equiv H_{uz}$. In this case the variational derivative of
the determinant $\frac{\delta \Gamma}{\delta A_{i}^{a}(x)}$ can be
computed to give \cite{L}
\begin{equation}
\delta \Gamma = - \frac{2N \mu}{\pi^{3/2}}
\int tr[ (H^{\dagger} H)^{-1} \delta(H H^{\dagger}) \partial ((H^ 
{\dagger} H)^{-1} \bar{\partial} H^{\dagger} H)]
\end{equation}
This expression integrates to a $3$ dimensional generalization of the  
WZW action
\begin{equation}
\Gamma = - \frac{2N \mu}{\pi^{3/2}}\left[\frac{1}{2} \int du d^2z\ \tr 
(\partial(H^{\dagger} H)
\bar{\partial}(H^{\dagger} H)^{-1})
+ \frac{i}{12} \int du \int_{B_u} \tr [(H^{\dagger} H)^{-1} d(H^ 
{\dagger} H)]^3\right]
\end{equation}
where $B_u$ denotes a $3$-ball bounding the plane $u = const$.

In this form the parallel with the KN formalism
\cite{knair} becomes very striking. The $2+1$ dimensional results
developed in \cite{knair} can be thus viewed as a natural reduction of
the $3+1$ dimensional formulation in the semi-complex basis.
In particular, the Hamiltonian in the semi-complex basis is \cite{4dpapers} (focusing 
on the $J_{z\bar{z}}$ dependence)
\be
{\cal H} \sim \frac{g^2 N \mu}{ {2\pi}^{3/2}}\left\{ \int du d^{2}z\  J_{z 
\bar z}\frac{\delta}{\delta J_{z\bar z}}+
\int du \int d^{2}w_1 d^2{w_2}\ \Omega(w_1, w_2) \frac{\delta}{\delta  
J_{z\bar z}(w_1)}\frac{\delta}{\delta J_{z\bar z}(w_2)}
+\ldots\right\}
\ee
where $\Omega$ is essentially fixed by the two-point function of the  
WZW model as in the $2+1$ dimensional context \cite{knair, prl}.

\section{Vacuum wave-functional: general discussion}

In this section, we make some general comments about the vacuum
wave-functional in $3+1$ dimensional Yang Mills theory in view of the  
already
discussed parallel between pure gauge theories in $2+1$ and $3+1$
dimensions. First, on general grounds one knows that the vacuum
wave-functional $\Psi_0$ is gauge invariant, parity even (in the  
absence of
$\theta$ terms) and most importantly should be a strictly positive
functional. This can be written as
\be
\Psi_0 = e^P
\ee
where $P$ is a functional of the `position' variables. Second, we also
know that the vacuum wave-functional can be formally written as a path
integral on half space-time as
\begin{equation}
\Psi_0[\tilde{A}] = \int_{A(t=0) = \tilde{A}} DA\
e^{\frac{1}{2g^2} \int_{-\infty}^{0} dt \int d^3x\ \tr (F_{\mu \nu}  
F^{\mu \nu}) (t,x)}
\end{equation}
where we integrate over all fields satisfying the boundary condition
$A(0,x) = \tilde{A}(x)$ and the integral is computed in the $A_0=0$
gauge. This implies that $P$ is a gauge invariant sum over connected
Feynman diagrams and in the large $N$ limit is a non-local functional  
involving only a {\it single trace} over gauge indices.

A physical mass scale $m$ must emerge in this theory. In fact, in  
order to write a wave-functional that involves arbitrary number of  
derivatives acting on fields, a scale must be introduced by hand.  
Presumably then, physical mass scales would be determined self- 
consistently. Furthermore, if the $3+1$ theory follows the $2+1$  
theory, we can suppose that in the IR, the mass scale enforces  
locality of $P$ in this limit. Ultimately, this question can only be  
answered for sure by computing physical quantities (such as the  
Wilson loop expectation value). Given this assumption, we can  
formally develop the non-local functional $P$ in
terms of local operators. By the previous argument, this expression
involves only local, single trace gauge invariant functionals. Such  
operators are naturally
labeled by two integers $n$, $k$ which count the number of laplacians
and powers of magnetic field insertions respectively. We can write this
expansion, in parallel with $2+1$, symbolically as
\begin{equation}
P = \frac{m^3}{2g^2} \sum_{k,n} \int \tr\left[\left(\frac{F}{m^2} 
\right)^k \left(\frac{\Delta}{m^2}\right)^n\right]
\end{equation}

We will advocate, as in $2+1$, that it is useful to take $P$ to have  
a certain quadratic form. In particular, when written in terms of the  
gauge invariant variables, we will consider as an ansatz
\begin{equation}
\Psi_0 = \exp\left(  \frac{1}{g^2 m} \int \tr\ B_{i} \ K(L/m^2) B_ 
{i}  +\ldots\right).
\end{equation}
where $K$ is a kernel constructed out of `holomorphic' covariant  
derivatives scaled by $m$, but does not contain the curvature $B_i$.  
The kernel $K$ should be determined (although we do not do so here)  
consistent with gauge, `holomorphic' and spacetime symmetries. Most  
importantly, it should be determined such that the wave-functional is  
{\it normalizable}. In $2+1$, such a kernel was found, and it seems  
reasonable to suppose that this could be repeated in $3+1$, given the  
similarity of the formalism.

We note, independent of these remarks, that
$P$ should have a specific asymptotic form in the far UV --- the  
theory is free in that limit, and thus the vacuum
wave-functional should be appropriate to free gluons. In this limit, we
should find
\be
P_{UV} = \frac{1}{g^2}\int_k  \tr\
B_i(-k) \frac{1}{\sqrt{\vec k^2}} B_i(k).
\ee
This is determined essentially by dimensional
analysis and is consistent with the transversality of gluons. We note  
that the mass scale $m$ introduced to define the kernel, disappears  
in this limit. It is a straightforward exercise to demonstrate that  
the Schr\"odinger formalism does indeed reproduce this perturbative  
result.

The UV behavior of the wave-functional is required but does not  
address the question of normalizability. Also, to be consistent with  
confinement, (in the sense of no phase transition occurring between  
UV and IR) the kernel $K$ should behave, thought of as a function of  
momentum, in a smooth way, and as we have discussed, take a certain  
simple form in the far IR. In particular, we are supposing here that  
this form is
\be\label{PIR}
P_{IR}= \frac{1}{2g^2 m}\int_k \tr\ B_i(-k) B_i(k).
\ee
Another way to motivate this form would be to compute how the  
Hamiltonian acts on this operator, and build the Schr\"odinger  
equation. Using the `holomorphic' regularization \cite{L} with a  
regulator scale $\mu$ and in the context of real variables, one finds
\be\label{Lact}
g^2\mathcal{T}\cdot \int B_{i}^{a}B_{i}^{a} = 2M \int B_{i}^{a}B_{i}^{a}
\ee
where
\be
2M=\frac{g^2N}{2(2\pi)^{3/2}}\ \mu.
\ee
This is reminiscent of the behavior in $2+1$, and is consistent with  
the kinetic energy acting to `count' derivatives.
Perhaps the simplest way to understand this result is to resort to the
semi-complex basis in which the intuition gained in $2+1$ dimensions
becomes useful. In that basis, there is a homogeneous part in the
kinetic term (as in our toy example) which acts as a dimension operator.
\be
{\cal H}\sim M \int J_{z\bar z}\frac{\delta}{\delta J_{z\bar z}}+\ldots
\ee
where $M$ is an effective mass scale. This is a first order operator.  
The rest of the
kinetic term, a second order operator, acts to properly normal order the
operators in the regularized calculation as well as provide the
necessary invariant counterterms so that the final result is indeed
holomorphic invariant.

Note though that the difference between $2+1$ and $3+1$ is in the  
power-counting. In $2+1$, $M$ is naturally given by the
coupling constant due to its dimensionality. In $3+1$ the coupling
constant is dimensionless, and $M$ is generated dynamically from the
regularization of $\mathcal{T}$, and appears for obvious dimensional  
reasons! It becomes clear though that this result is regulator  
dependent, and what happens in the continuum limit is not clear.

The result (\ref{Lact}) is consistent with (\ref{PIR}), relating $M$  
to $m$. The IR vacuum wave-functional provides a probability measure $ 
\Psi_0^* \Psi_0$ equivalent to the partition function of the  
Euclidean three-dimensional Yang-Mills theory with an effective Yang- 
Mills coupling $g_{3D}^2 \equiv m g^2$. One could use this to compute  
the expectation value of a large Wilson loop and deduce the area law  
behavior. Presumably this would mean that the square root of the  
string tension scales with $m$. This remark ties the scale introduced  
into the vacuum wave-functional to this physical parameter.

Let us remark further on the generalized Gaussian form of the vacuum  
wave-functional. In particular, the
neglect of higher order terms in the exponent of the wave-functional
needs to be justified. We want to be perfectly clear that the validity
of the generalized Gaussian form for the vacuum wave-functional cannot
be established by appealing only to standard large $N$ simplifications.
The large $N$ limit only selects single trace expressions in the ansatz
for the wave-functional. Thus we also do not expect that this
wave-functional is exact; if this were true, $B_i$  would indeed
represent the right variables appropriate to the large $N$ limit.
Nevertheless, as in the $2+1$ dimensional case \cite{prl}, these local
gauge invariant variables are the correct constituent degrees of
freedom, even though they do not appear as physical asymptotic states!
The physical states (an infinite number of glueballs, or non-local gauge
invariant variables, such as Wilson loops) can be built out of these
local degrees of freedom, so that all expected predictions based on the
large $N$ counting (factorization, suppression of vertices etc) are
fulfilled.

The neglect of the higher order terms in $B_i$ suggests that there is a
second expansion parameter at work, involving a new length scale. One
such candidate scale is the size of glueballs. The approximation
employed here amounts to considering only `free' glueballs, which are
point-like and non-interacting. This is consistent with the large $N$
picture where one expects that all glueball interactions are suppressed
by powers of $1/N$. This wave-functional has the form of a ``generalized
coherent state'' appropriate to large $N$ \cite{collective}. One way to
intuitively motivate the Gaussian form of the wave-functional is to
think of $B_i$ as the relevant local probes of real physical states, but
not as actual asymptotic states! Then apart from the rank of the gauge
group $N$, there should exist another expansion parameter, which is
related to the size of the glueballs. The quadratic term in the wave
functional should be then interpreted as the leading term in the
expansion in the inverse of that effective glueball size. This would be
very reminiscent of the $\alpha'$ expansion in string theory.

As we mentioned above the confinement should be implied by the
normalizability of the wave-functional, as in the $2+1$ dimensional
counterpart \cite{prl}. Note that the constituent $B_i$ should not  
appear as
an asymptotic state. Also, we further expect the glueball  
constituents to be
``seeds'' for constituent quarks once the fermionic degrees of freedom
are included.

Furthermore we also want to reiterate that the ``QCD scale'' as given
by $m$ should be self-consistently determined: on one side $m$ is needed
in the wave-functional for dimensional reasons, and thus it sets the  
mass scale for the
spectrum, and on the other hand, the
square root of the string tension is given in terms of $m$ up to a
multiplicative numerical constant. Note that the cut-off $\mu$ appears
both in the expression for the gap and the string tension, and should
consistently cancel in the ratio (assuming they both persist in the  
continuum). Also, in the limit of small coupling, the ratio of
the string tension and the scale $m$ (or equivalently the
cut-off $\mu$) should only depend on the dimensionless coupling $g^2 N$,
and thus given the fact that the wave-functional describes a free theory
in the UV, is then corrected logarithmically, as implied by asymptotic
freedom.


Finally, one of the amazing features of large $N$ lattice simulations
\cite{T} is that the actual numerical values for the ratio of masses and
the square root of the string tension are of the same order of
magnitude, both in $2+1$ and $3+1$ dimensions. The actual numbers are
very close, up to $15\%$ to $20\%$. We think that this is
evidence that the large $N$ Yang Mills theories in $2+1$ and $3+1$
theories are ``close'' in the sense of the respective continuum limits.
We think that this is not a coincidence in view of the formalism we have
discussed.

\section{Conclusion}

In this short programmatic paper we have discussed an
analytic approach towards the solution of pure Yang-Mills theory in
$3+1$ dimensional spacetime. Our approach is based on the use
of local gauge invariant corner variables in the Schr\"odinger
representation and the large $N$ limit. In particular, within this
approach one finds unexpected parallels between pure Yang-Mills theory
in $2+1$ and $3+1$ dimensions.

There are obviously many important questions to be addressed. Perhaps
the most important is to determine a definite form of the kernel in the
quadratic wave-functional in parallel with the successful discussion of
the $2+1$ dimensional situation \cite{prl} which did lead to the
computation of the large $N$ glueball spectra/Regge trajectories. The
large $N$ lattice results are already available \cite{T}.

There are of course many other obvious questions, both in the $2+1$ and
$3+1$ dimensional contexts: the inclusion of fermions and the
computation of the meson (and baryon) masses in the large N limit; the
development of the covariant approach; the elucidation of the role of
topology; the proof of confinement and many others. We plan to address
some of these questions in \cite{4dpapers}.

\vskip 1cm

{\bf \Large Acknowledgements}

We would like to thank A. Yelnikov for many discussions and
collaboration. We would also like to thank R. Brout, K. Gawedzki, V. P.
Nair, M. Rozali, L. Smolin and T.Takeuchi for interesting conversations
regarding the material presented in this paper. We also thank I. Bars
for e-mail communications concerning his work. {\small RGL}
was supported in part by the U.S. Department of Energy under contract
DE-FG02-91ER40709. {\small DM} is supported in part by the U.S.
Department of Energy under contract DE-FG05-92ER40677. Many thanks to
Perimeter Institute for hospitality.

\end{document}